\documentclass[onecolumn,10pt]{article}
\usepackage{amsmath}
\usepackage{amsfonts}
\usepackage{amssymb}
\usepackage{amsthm}
\usepackage{times}
\usepackage{cancel}
\usepackage{graphics}

\newcommand{\hil}[1]{\mbox{$\mathcal{#1}$}}

\newcommand{\ket}[1]{| #1 \rangle}

\usepackage{tikz}

\title{\textbf{Contextuality and Nonlocality\\ in `No Signaling' Theories}}
\author{Jeffrey Bub\thanks{email address: jbub@umd.edu} \\ \small \textit{Philosophy Department and Institute for Physical Science and Technology}\\  \small \textit{University of Maryland, College Park, MD 20742, USA}\\ Allen Stairs \thanks{email address: stairs@umd.edu}\\ \small \textit{Philosophy Department, University of Maryland, College Park, MD 20742, USA}  }
\date{}

\normalsize
\begin{document}

\maketitle

\begin{abstract}

\end{abstract} 
We define a family of `no signaling' bipartite boxes with arbitrary inputs and binary outputs, and with a range of marginal probabilities. The defining correlations are motivated by the Klyachko version of the Kochen-Specker theorem, so we call these boxes Kochen-Specker-Klyachko boxes or, briefly, KS-boxes. The marginals  cover a variety of cases, from those that can be simulated classically to the superquantum correlations that saturate the Clauser-Horne-Shimony-Holt inequality, when  the KS-box is a generalized PR-box (hence a vertex of the `no signaling' polytope). We show that for certain marginal probabilities a KS-box is classical with respect to nonlocality as measured by the Clauser-Horne-Shimony-Holt correlation, i.e.,  no better than shared randomness as a resource in simulating a PR-box, even though such KS-boxes cannot be perfectly simulated by classical or quantum resources for all inputs. We comment on the significance of these results for  contextuality and nonlocality in `no signaling' theories.

\bigskip


\section{Introduction}

Recently, there has been considerable interest in studying correlations between separated systems in `no signaling' theories, which include subquantum (e.g., classical), quantum, and superquantum theories. The primary foundational aim is to characterize quantum mechanics, i.e., to identify physical principles that distinguish quantum mechanics from other theories that satisfy a `no signaling' principle (see below). 

From an information-theoretic standpoint, a classical state space has the structure of a simplex. An $n$-simplex is a particular sort of convex  set: a convex polytope generated by $n+1$ vertices that are not confined to any $(n-1)$-dimensional subspace (e.g., a triangle as opposed to a rectangle). The simplest classical state space is the 1-bit space (1-simplex), consisting of two pure or extremal deterministic states, $\mathbf{0} = \left ( \begin{array}{c}1 \\ 0 \end{array} \right )$ and $\mathbf{1} = \left ( \begin{array}{c}0 \\ 1 \end{array} \right )$, represented by the vertices of the simplex, with mixtures---convex combinations of pure states---represented by the line segment between the two vertices: $\mathbf{p} = p\,\mathbf{0} + (1-p)\,\mathbf{1}$, for $0 \leq p \leq 1$. A simplex has the rather special property that a mixed state can be represented in one and only one way as a mixture of pure states, the vertices of the simplex. No other state space has this feature: if the state space is not a simplex, the representation of mixed states as convex combinations of pure states is not unique. The state space of classical mechanics is an infinite-dimensional simplex, where the pure states are all deterministic states, with enough structure to support  transformations acting on the vertices that include the canonical transformations generated by Hamiltonians. 

The simplest quantum system is the qubit, whose state space as a convex set has the structure of a sphere (the Bloch sphere), which is not a simplex. The non-unique decomposition of mixtures into pure states underlies the impossibility of a universal cloning operation for pure states in  nonclassical theories or, more generally, the impossibility of a universal broadcasting operation for an arbitrary set of states, and the monogamy of nonclassical correlations, which are generic features of non-simplex theories \cite{Masanes2006}. 

The space of `no signaling' probability distributions is a convex polytope that is not a simplex (see \cite{JonesMasanes2005}, \cite{BLMPP2005}, \cite{BarrettPironio2005}). Some of these vertices are non-deterministic Popescu-Rohrlich (PR) boxes \cite{PopescuRohrlich94}, or generalizations of PR-boxes. A PR-box is a hypothetical device or nonlocal information channel that is more nonlocal than quantum mechanics, in the sense that the correlations between outputs of the box for given inputs violate the Tsirelson bound \cite{Tsirelson1980}.  A PR-box is defined as follows: there are two inputs, $x \in \{0,1\}$ and $y\in \{0,1\}$, and two outputs, $a\in \{0,1\}$ and $b\in \{0,1\}$. The box is bipartite and nonlocal in the sense that the $x$-input and $a$-ouput can be separated from the $y$-input and $b$-output by any distance without altering the correlations. For convenience, we can think of the  $x$-input as controlled by Alice, who monitors the $a$-ouput, and the $y$-input as controlled by Bob, who monitors the $b$-output. Alice's and Bob's inputs and outputs are then required to be correlated according to:
\begin{equation}
a\oplus b = x.y \label{eqn:PRbox}
\end{equation}
where $\oplus$ is addition mod 2, i.e., 
\begin{itemize}
\item[(i)] same outputs (i.e., 00 or 11) if the inputs are 00 or 01 or 10 
\item[(ii)] different outputs (i.e., 01 or 10) if  the inputs are 11
\end{itemize}

The `no signaling' condition is a requirement on the marginal probabilities: the marginal probability of Alice's outputs do not depend on Bob's input,  i.e., Alice cannot tell what Bob's input was by looking at the statistics of her outputs, and conversely. Formally:
\begin{eqnarray}
\sum_{b\in\{0,1\}}p(a,b|x,y) = p(a|x), \, a, x, y \in\{0,1\} \\
\sum_{a\in\{0,1\}}p(a,b|x,y) = p(b|y), \, b, x, y \in\{0,1\}
\end{eqnarray}
The correlations (\ref{eqn:PRbox}) together with the `no signaling' condition entail that the marginals are equal to 1/2 for all inputs $x, y\in\{0,1\}$ and all outputs $a,b\in\{0,1\}$:
\begin{equation}
p(a=0|x) = p(a=1|x) = p(b=0|y) = p(b=1|y) = 1/2
\end{equation}

A PR-box can be defined equivalently in terms of the joint probabilities for all inputs and all outputs, as in Table 1. For bipartite probability distributions, with two input values and two output values, the vertices of the `no signaling' polytope are all PR-boxes (differing only with respect to permutations of the input values and/or output values) or deterministic boxes.
\begin{table}[h!]
\begin{center}
\begin{tabular}{|ll||ll|ll|} \hline
   &$x$&$0$ & &$1$&\\
   $y$&&&&&\\\hline\hline
  $0$ &&$p(00|00) = 1/2$&$ p(10|00 = 0) = 0$  & $p(00|10) = 1/2$&$ p(10|10) = 0$     \\
   &&$p(01|00) = 0$&$p(11|00) = 1/2$  & $p(01|1)=0$&$ p(11|10) = 1/2$  \\\hline
   $1$&&$p(00|01)=1/2$&$ p(10|01)=0$  & $p(00|11=0$&$ p(10|11)=1/2$   \\
  &&$p(01|01)=0$&$ p(11|01)=1/2$  & $p(01|11)=1/2$&$ p(11|11)=0$   \\\hline
\end{tabular}
\end{center}
 \caption{Joint probabilities for the PR-box}
\end{table}

Consider the problem of simulating a PR-box: how close can Alice and Bob come to simulating the correlations of a PR-box for random inputs if they are limited to certain resources? In units where $a = \pm 1, b = \pm 1$,
\begin{equation}
\langle 00\rangle = p(\mbox{same output}|00) - p(\mbox{different output}|00)
\end{equation}
so:
\begin{eqnarray}
p(\mbox{same output}|00) & = & \frac{1 +  \langle 00\rangle}{2} \\
p(\mbox{different output}|00) & = & \frac{1-\langle 00\rangle}{2}
\end{eqnarray}
and similarly for input pairs 01, 10, 11. It follows that the probability of a successful simulation is given by:
\begin{eqnarray}
\mbox{prob(successful sim)} & = & \frac{1}{4}(p(\mbox{same output}|00) + p(\mbox{same output}|01) + \nonumber \\
& &  p(\mbox{same output}|10) + p(\mbox{different output}|11)) \\
& = & \frac{K}{8} + \frac{1}{2}
\end{eqnarray}
where $K = \langle 00\rangle + \langle 01\rangle + \langle 10\rangle - \langle 11\rangle$ is the Clauser-Horne-Shimony-Holt (CHSH) correlation. 

Bell's locality argument \cite{BellEPR} in the CHSH version \cite{CHSH} shows that if Alice and Bob are limited to classical resources, i.e., if they are required to reproduce the correlations on the basis of shared randomness or common causes established before they separate (after which no communication is allowed), then $K_{C} \leq 2$, so  the optimal probability of success is 3/4. If Alice and Bob are allowed to base their strategy on shared entangled states prepared before they separate, then the Tsirelson inequality requires that $K_{Q} \leq 2\sqrt{2}$, so the optimal probability of success limited by quantum resources is approximately .85. For the PR-box, K = 4, so the probability of success is, of course, 1. 

It is easy to show that the correlations of a PR-box are monogamous and that the pure states, defined as above, cannot be cloned \cite{Masanes2006}.  In a recent paper \cite{SkrzypczykBrunnerPopescu2008}, the authors introduce a dynamics for PR-boxes and show that the Tsirelson bound defines the limit of  nonlocality swapping for noisy PR-boxes. This is a very remarkable result about the nonlocality of nonclassical `no signaling' theories. 

Before Schr\"{o}dinger  \cite[p. 555]{Schr1} characterized nonlocal entanglement as  `\textit{the} characteristic trait of
quantum mechanics, the one that enforces its entire departure from
classical lines of thought,' another feature of quantum mechanics, emphasized by Bohr, was generally regarded as the distinguishing feature of quantum systems:  the apparent dependence of measured values on the local experimental context. This contextuality is exhibited in various 
ways---noncommutativity, the uncertainty principle, the impossibility of assigning values to all observables of a quantum system simultaneously, while requiring the functional relationships between observables to hold for the corresponding values (so, e.g., the value assigned to the square of an observable should be the square of the value assigned to the observable)---but for our purposes here the relevant result is the theorem by Kochen and Specker \cite{KochenSpecker}. 

Kochen and Specker identified a finite noncommuting set of 1-dimensional projection operators on a 3-dimensional Hilbert space, in which an individual projection operator can belong to different orthogonal triples of projection operators representing different bases or contexts, such that no assignment of 0 and 1 values to the projection operators is possible that is both (i) \emph{noncontextual} (i.e., each projection operator is assigned one and only one value, independent of context), and (ii) \emph{respects the orthogonality relations }(so that the assignment of the value 1 to a 1-dimensional  projection operator $P$ requires the assignment of 0 to any projection operator orthogonal to $P$). A quantum system associated with a 3-dimensional Hilbert space is only required to produce a value for an observable represented by a 1-dimensional projection operator $P$ with respect to the context defined by $P$ and its orthogonal complement $P^{\perp}$ in a nonmaximal measurement, or with respect to a context defined by a particular orthogonal triple of projection operators in a maximal measurement. Unlike the situation in classical mechanics, different maximal measurement contexts for a quantum system are exclusive, or `incompatible' in Bohr's terminology: they cannot all be embedded into one context. In this sense, measurement in quantum mechanics is contextual, and the distribution of measurement outcomes for a quantum state cannot be simulated by a noncontextual assignment of values to all observables, or even to certain finite sets of observables, by the Kochen-Specker theorem. 

Note that the Kochen-Specker result does not justify the claim  that the outcome of a measurement of an observable would have been different if the observable had been measured with respect to  a different context. This is a counterfactual statement concerning an unperformed measurement, and---as Asher Peres was fond of repeating---unperformed measurements have no results: there is, in principle, no way to check this claim.  Note also that the contextuality of individual measurement outcomes is masked by the statistics, which is noncontextual:  the probability that a measurement of an observable corresponding to a projection operator $P$ yields the value 1 in a quantum state $\ket{\psi}$ is the same, irrespective of the measurement context, i.e., irrespective of what other projection operators are measured together with $P$ in the state $\ket{\psi}$. Similarly, the effect of nonlocality in quantum mechanics is not directly represented in the statistics: there is no violation of the `no signaling' principle---Alice's statistics is unaffected by Bob's measurements. 

Locality in Bell's sense is a probabilistic noncontextuality constraint with respect to \emph{remote} contexts. Specifically, in terms of the inputs and outputs of a PR-box, locality is the requirement that (I) the probability of a given output, $a$ of $x$, \emph{conditional on a shared random variable for the two inputs $x$ and $y$,} is independent of the remote $y$-context, and also (II) independent of the outcome $b$ for a given remote $y$-context (and conversely). Note that (I) is not the same as the `no signaling' condition, which is (I) without the qualification `conditional on a shared random variable.' That is, the `no signaling' condition refers to `surface probabilities,' while the condition (I) refers to `hidden probabilities' (to use a terminology due to van Fraassen \cite{Fraassen1982}). Shimony \cite{Shimony2006} calls conditions (I) and (II)  `parameter independence' and `outcome independence,' respectively.\footnote{This formulation of Bell's locality condition as the conjunction of two  independent conditions was first proposed by Jarrett  \cite{Jarrett1984}, who called (I) `locality' and (II) `completeness.'}

In the following, we define a family of bipartite boxes with $n$ possible input values instead of two, i.e., $x \in \{1, \ldots, n\}, y \in \{1, \ldots, n\}$, and binary outputs $a \in \{0,1\}, b \in \{0,1\}$, which allows the consideration of a  range of nonlocal contexts defined by pairs of inputs to the box. In a quantum simulation based on a strategy that exploits shared entangled states to reproduce the correlations, the inputs are associated with measurements of specific observables, and the nonlocal  box contexts can be associated with different local measurement contexts that share a common element corresponding to an input value. The correlational constraints are motivated by a version of the Kochen-Specker theorem due to Klyachko \cite{Klyachko2002,Klyachko2007}, so we call such boxes Kochen-Specker-Klyachko boxes or, briefly, KS-boxes.

The family of KS-boxes is parametrized by the marginal probability $p$ for the output 1, where $0 \leq p \leq 1/2$. The marginals cover a range of cases, from those that can be simulated classically to the superquantum correlations that saturate the Clauser-Horne-Shimony-Holt inequality, when  the KS-box is a generalization of a PR-box and hence a vertex of the `no signaling' polytope. For certain marginal probabilities, a KS-box can display correlations that are no more nonlocal than classical correlations, as measured by the CHSH correlation, even though a perfect simulation of the correlations for all inputs with classical or quantum resources is impossible.

We sketch Klyachko's version of the Kochen-Specker theorem in \S 2. The defining correlations of the KS-box are set out in \S 3, where we consider the issue of simulating a KS-box with classical or quantum resources. In \S 4, we consider simulating a PR-box with a KS-box and show that, for a marginal probability $p = 1/3$, a KS-box is no better than shared randomness as a resource in simulating the correlations of a PR-box, even though  the KS-box cannot be perfectly simulated by classical or quantum resources for all inputs. In \S 5, we drop the marginal constraint and consider the behavior of a KS-box for all marginal probabilities $0 \leq p \leq 1/2$. We conclude in \S 6 with some remarks commenting on the significance of these results for contextuality and nonlocality in `no signaling' theories. 

\section{Klyachko's Version of the Kochen-Specker Theorem}

Consider a unit sphere and imagine a circle $\Sigma_{1}$ on the equator of the sphere with an inscribed pentagon and pentagram, with the vertices of the pentagram labelled in order 1, 2, 3, 4, 5 (see Fig. 1).\footnote{The following formulation of Klyachko's proof owes much to a discussion with Ben Toner and differs from the analysis in \cite{Klyachko2002,Klyachko2007}.}
Note that the angle subtended at the center $O$ by adjacent vertices of the pentagram defining an edge (e.g., 1 and 2) is $\theta = 4\pi/5$, which is greater than $\pi/2$. It follows that if the radii linking O to the vertices are pulled upwards towards the north pole of the sphere, the circle with the inscribed pentagon and pentagram will move up on the sphere towards the north pole. Since $\theta = 0$ when the radii point to the north pole (and the circle vanishes), $\theta$ must  pass through $\pi/2$ before the radii point to the north pole, which means that it is possible to draw a circle $\Sigma_{2}$ with an inscribed pentagon and pentagram on the sphere at some point between the equator and the north pole, \emph{such that the angle subtended at $O$ by an edge of the pentagram is $\pi/2$}.  We label the centre of this circle $P$ (see Fig. 2; note that the line OP is orthogonal to the circle $\Sigma_{2}$ and is not in the plane of the pentagram). 

 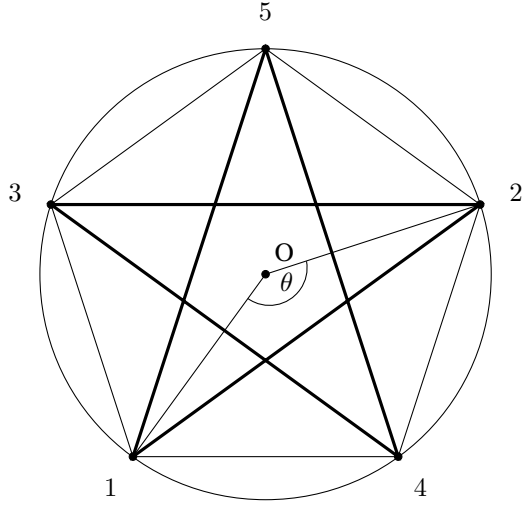
\begin{figure}[!ht]
    \begin{picture}(300,180)(-65,0)
\begin{tikzpicture}
\tikzstyle vertex=[circle,draw,fill=black,inner sep=1pt]
\path (0,0) coordinate (O);
\path (3*72+18:3cm) coordinate (P1);
\path (18:3cm) coordinate (P2);
\path (2*72+18:3cm) coordinate (P3);
\path (4*72+18:3cm) coordinate (P4);
\path (72+18:3cm) coordinate (P5);
\path (3*72+18:.4cm) coordinate (Q);
\node at (.25,.3)  {O};
\node at (-20:.3cm)  {$\theta$};
\path (3*72+18:3.5cm)
node {$1$};
\path (18:3.5cm) 
node {$2$};
\path (2*72+18:3.5cm) 
node {$3$};
\path (4*72+18:3.5cm)
node {$4$};
\path (72+18:3.5cm) 
node {$5$};
\draw[very thick] (P1) -- (P2) -- (P3) -- (P4) -- (P5) -- cycle;
\draw (P1) -- (P4) (P4) -- (P2)
(P2) -- (P5) (P5) -- (P3)
(P3) -- (P1);	
\draw (P1) -- (O) (O) -- (P2);
\draw (Q) arc (-125:10:.5cm);
\draw (0,0) circle (3cm);
\node[vertex] at (0,0) {};
\node[vertex] at (P1) {};
\node[vertex] at (P2) {};
\node[vertex] at (P3) {};
\node[vertex] at (P4) {};
\node[vertex] at (P5) {};
\end{tikzpicture}
    \label{fig1}
\end{picture}
\caption{Circle $\Sigma_{1}$ with inscribed pentagram}
 \end{figure}

One can therefore define five orthogonal triples of vectors, i.e., five bases in a 3-dimensional Hilbert space $\hil{H}_{3}$, representing five different measurement contexts:
\[ \begin{array}{lll}
\ket{1},&\ket{2},&\ket{v} \\
\ket{2},& \ket{3}, & \ket{w}  \\
\ket{3}, & \ket{4}, & \ket{x} \\
\ket{4}, & \ket{5}, & \ket{y}  \\
\ket{5}, & \ket{1}, & \ket{z}
\end{array} \]
Here $\ket{v}$ is orthogonal to $\ket{1}$ and $\ket{2}$, etc. Note that each vector $\ket{1},\ket{2}, \ket{3}, \ket{4}, \ket{5}$ belongs to two different contexts. The vectors $\ket{u}, \ket{v}, \ket{x}, \ket{y}, \ket{z}$ play no role in the following analysis, and we can take a context as defined by an edge of the pentagram in the circle $\Sigma_{2}$.

Consider, now, assigning 0's and 1's to all the vertices of the pentagram in $\Sigma_{2}$ noncontextually (i.e., each vertex is assigned a value independently of the edge to which it belongs), in such a way as to satisfy the orthogonality constraint that at most one 1 can be assigned to the two vertices of an edge. It is  obvious by inspection that the orthogonality constraint can be satisfied noncontextually by assignments of zero 1's, one 1, or two 1's (but not by three 1's, four 1's, or five 1's). Call such assignments `charts.' We say that the constraints can be satisfied by charts of type $C_{0}, C_{1}, C_{2}$. It follows that for such charts, where $v(i)$ is the value assigned to the vertex $i$:
\begin{equation}
\sum_{i=1}^{5}v(i) \leq 2
\end{equation}

If we label the possible charts with a hidden variable $\lambda \in \Lambda$, and average over $\Lambda$, then the probability of a vertex being assigned the value 1 is given by:
\begin{equation}
p(v(i) =1) = \sum_{\Lambda} v(i|\lambda)p(\lambda)
\end{equation}
so:
\begin{eqnarray}
\sum_{i=1}^{5}p(v(i)=1) & = & \sum_{i=1}^{5}\sum_{\Lambda} v(i|\lambda)p(\lambda)\nonumber \\
& = & \sum_{\Lambda} (\sum_{i=1}^{5} v(i|\lambda))p(\lambda) \leq 2 \label{Klyachko}
\end{eqnarray}

We have shown that the sum of the probabilities assigned to the vertices of the pentagram on the circle $\Sigma_{2}$ must be less than or equal to 2, if the selection of a vertex (denoted by the assignment of 1) is made noncontextually in such a way as to satisfy the orthogonality constraint. Note that this Klyachko inequality follows without any assumption about the relative weighting of the charts. 

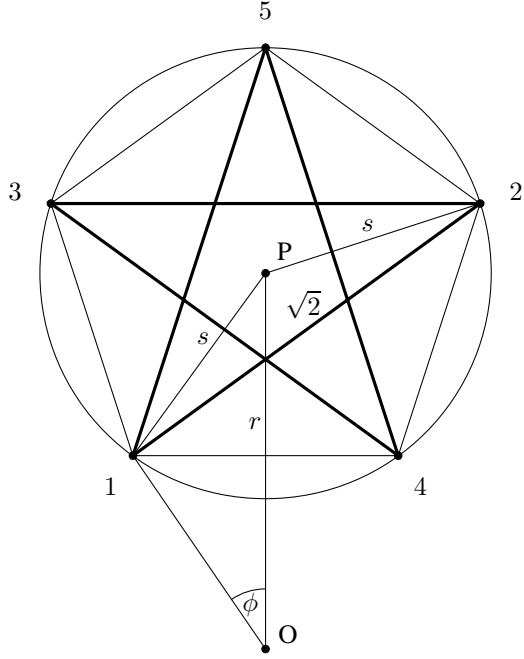
\begin{figure}[!ht]
    \begin{picture}(300,250)(-65,0)
\begin{tikzpicture}
\tikzstyle vertex=[circle,draw,fill=black,inner sep=1pt]
\path (0,0) coordinate (P);
\path (0,-5) coordinate (O);
\path (3*72+18:3cm) coordinate (P1);
\path (18:3cm) coordinate (P2);
\path (2*72+18:3cm) coordinate (P3);
\path (4*72+18:3cm) coordinate (P4);
\path (72+18:3cm) coordinate (P5);
\path (0,-4.2) coordinate (Q);
\node at (.25,.3)  {P};
\node at (.3,-4.8)  {O};
\node at (-.2,-4.4)  {$\phi$};
\path (3*72+18:3.5cm)
node {$1$};
\path (3*72+10:1.2cm)
node {$s$};
\path (18:3.5cm) 
node {$2$};
\path (25:1.5cm) 
node {$s$};
\path (-94:2cm) 
node {$r$};
\path (-40:.65cm) 
node {$\sqrt{2}$};
\path (2*72+18:3.5cm) 
node {$3$};
\path (4*72+18:3.5cm)
node {$4$};
\path (72+18:3.5cm) 
node {$5$};
\draw[very thick] (P1) -- (P2) -- (P3) -- (P4) -- (P5) -- cycle;
\draw (P1) -- (P4) (P4) -- (P2)
(P2) -- (P5) (P5) -- (P3)
(P3) -- (P1);	
\draw (O) -- (P) (O) -- (P1) (P) -- (P2) (P) -- (P1); 
\draw (Q) arc (90:125:.8cm);
\draw (0,0) circle (3cm);
\node[vertex] at (0,0) {};
\node[vertex] at (P1) {};
\node[vertex] at (P2) {};
\node[vertex] at (P3) {};
\node[vertex] at (P4) {};
\node[vertex] at (P5) {};
\node[vertex] at (O) {};

\end{tikzpicture}

\end{picture}
\caption{Circle $\Sigma_{2}$ with inscribed pentagram}
    \label{fig2}
\end{figure}

Now consider a quantum system in the state defined by a unit vector that passes through the north pole of the sphere. This vector passes through the point $P$ in the center of the circle $\Sigma_{2}$. Call this state $\ket{\psi}$. A simple geometric argument shows that if probabilities are assigned to the states or 1-dimensional projectors defined by the vertices of the pentagram on $\Sigma_{2}$ by the  state $\ket{\psi}$, then the sum of the probabilities is greater than 2!

To see this, note that the probability assigned to a vertex, say the vertex 1, is:
\begin{equation}
|\langle 1|\psi\rangle|^{2} = \cos^{2} \phi
\end{equation}
where $\ket{1}$ is the unit vector defined by the radius from O to the vertex 1. Since the lines from the center $O$ of the sphere to the vertices of an edge of the pentagram on $\Sigma_{2}$ are radii of length 1 subtending a right angle, each edge of the pentagram has length $\sqrt{2}$. The angle subtended at $P$ by the lines joining $P$ to the two vertices of an edge is $4\pi/5$, so the length, $s$, of the line joining $P$ to a vertex of the pentagram is:
\begin{equation}
s = \frac{1}{\sqrt{2}\cos \frac{\pi}{10}}
\end{equation}
Now, $\cos \phi = r$, where $r$ is the length of the line $OP$, and $r^{2} + s^{2} = 1$, so:
\begin{equation}
\cos^{2} \phi = r^{2} = 1-s^{2} =  \frac{\cos \frac{\pi}{5}}{1+\cos\frac{\pi}{5}} = 1/\sqrt{5}
\end{equation}
(because $\cos\pi/5 = \frac{1}{4}(1+\sqrt{5})$), and so:
\begin{equation}
\sum_{i=1}^{5} p(v(i) =1) = 5 \times 1/\sqrt{5} = \sqrt{5} > 2
\end{equation}

\section{The KS-Box}
\label{sec:KS box}

We define a KS-box as follows: The box has two inputs, $x, y \in\{1, \ldots, n\}$, and two outputs, $a, b \in \{0,1\}$. We call $n$ the dimension of the KS-box. As with a PR-box, we suppose that the $x$-input and $a$-output can be separated by any distance from the $y$-input and $b$-output without affecting the correlations, which are required to be:
\begin{enumerate}
\item[(i)] if $x = y$, then $a = b$
\item[(ii)] if $x \neq y$, then $a\cdot b = 0$
\end{enumerate}
That is, if the inputs are the same, the outputs are the same;   if the inputs are different, at least one output is 0 (i.e., both outputs cannot be 1). The marginal probabilities are required to satisfy the `no signaling' constraint. We shall consider KS-boxes with various marginals and show that they have different properties. We call a KS-box with a marginal probability of $p$ for the output 1 a KS$_{p}$-box.

In this section, we consider the problem of simulating a 5-dimensional KS-box with classical and quantum resources. For reasons that will become clear below (see the discussion in \S 6), $n=5$ is the smallest number of inputs for which the wider range of nonlocal contexts defined by input pairs $x, y$ precludes a perfect classical or quantum simulation for certain marginal probabilities. We also  initially require the marginal constraint $p = 1/3$, but we will eventually drop this constraint and consider the behavior of a KS$_{p}$-box for all marginal probabilities $0 \leq p \leq 1/2$ (while requiring, of course, `no signaling'). 

For convenience, we shall refer to the condition (ii)---if $x \neq y$ then $a.b=0$---as the `$\perp$' constraint, since it is motivated by the Kochen and Specker orthogonality condition requiring that an assignment of 1's and 0's to the 1-dimensional projection operators of a maximal context defined by a basis in Hilbert space should respect the orthogonality relations.  

Consider now the problem of simulating a $5$-dimensional KS$_{p}$-box, with $p = 1/3$,  with classical resources: to what extent can Alice and Bob simulate the correlations of the KS-box for random inputs if their only allowed resource is shared randomness? The requirement of perfect correlation if the inputs are the same forces local noncontextuality, i.e., Alice and Bob will have to base their strategy on shared charts selected by a shared random variable.

The pentagon edges and pentagram edges exhaust all possible input pairs $\{x,y\}$ for a 5-dimensional KS-box. For a marginal probability $p = 1/5$, a  perfect classical simulation of a 5-dimensional KS-box can be achieved with shared charts $C_{1}$, in which only a single vertex is assigned a 1. For marginals $p \leq 1/5$, a perfect classical simulation can be achieved if Alice and Bob mix the strategy for $p = 1/5$ and output 0 simultaneously and randomly for a certain fraction of agreed-upon rounds (i.e., before separation, they generate a random bit string with the appropriate probability of 0's, which they share, and they associate successive rounds of the simulation---successive input pairs---with elements of the string; when the shared bit is 0, they both output 0 independently of the input or, equivalently, they use chart $C_{0}$). In other words, they mix the above strategy with the strategy: `output 0 for any input,' with the appropriate mixture probabilities.  

Clearly, however, it is impossible to generate a marginal probability $p > 1/5$ without using charts $C_{2}$ as well. To satisfy the marginal constraint $p = 1/3$, Alice and Bob will have to adopt a strategy 
in which the output for a given input is based on either of the following two mixtures of shared charts, $M_{1}$ or $M_{2}$, selected by a shared random variable:
\begin{enumerate}
\item[$M_{1}$:] 2/3 $C_{2}$, 1/3 $C_{1}$ 
\item[$M_{2}$:] 5/6 $C_{2}$, 1/6 $C_{0}$ 
\end{enumerate}
or on mixtures of these two mixtures. 

We now observe that for charts $C_{2}$, the `$\perp$' constraint can be satisfied either for pentagon edges or for pentagram edges, but not both. See Fig. 3, where a chart $C_{2}$, indicated by the circled 0's and 1's, satisfies the `$\perp$' constraint for the pentagram edges. If the assigned value 1 is moved from the vertex 4 to the vertex 2, for example, the chart satisfies the `$\perp$' constraint for the pentagon edges, but violates the constraint for the pentagram edges. (For charts $C_{3}, C_{4}, C_{5}$, both pentagon edges and pentagram edges violate the `$\perp$' constraint.)

\begin{figure}[!ht]
    \begin{picture}(300,240)(-65,0)
\begin{tikzpicture}
\tikzstyle vertex=[circle,draw,fill=black,inner sep=1pt]
\path (0,0) coordinate (O);
\path (3*72+18:3cm) coordinate (P1);
\path (18:3cm) coordinate (P2);
\path (2*72+18:3cm) coordinate (P3);
\path (4*72+18:3cm) coordinate (P4);
\path (72+18:3cm) coordinate (P5);
\node at (.3,.2)  {O};
\path (3*72+18:3.5cm)
node {$1$};
\path (3*72+18:4.2cm)
node[draw,shape=circle] {$1$};
\path (18:3.5cm)
node {$2$};
\path (18:4.2cm) 
node[draw,shape=circle] {$0$};
\path (2*72+18:3.5cm) 
node {$3$};
\path (2*72+18:4.2cm) 
node[draw,shape=circle] {$0$};
\path (4*72+18:3.5cm)
node {$4$};
\path (4*72+18:4.2cm)
node[draw,shape=circle] {$1$};
\path (72+18:3.5cm) 
node {$5$};
\path (72+18:4.2cm) 
node[draw,shape=circle] {$0$};
\draw[very thick] (P1) -- (P2) -- (P3) -- (P4) -- (P5) -- cycle;
\draw (P1) -- (P4) (P4) -- (P2)
(P2) -- (P5) (P5) -- (P3)
(P3) -- (P1);	
\draw (0,0) circle (3cm);
\node[vertex] at (0,0) {};
\node[vertex] at (P1) {};
\node[vertex] at (P2) {};
\node[vertex] at (P3) {};
\node[vertex] at (P4) {};
\node[vertex] at (P5) {};
\end{tikzpicture}

\end{picture}
\caption{Chart $C_{2}$ satisfying `$\perp$' constraint for pentagram edges}
    \label{fig3}
\end{figure}
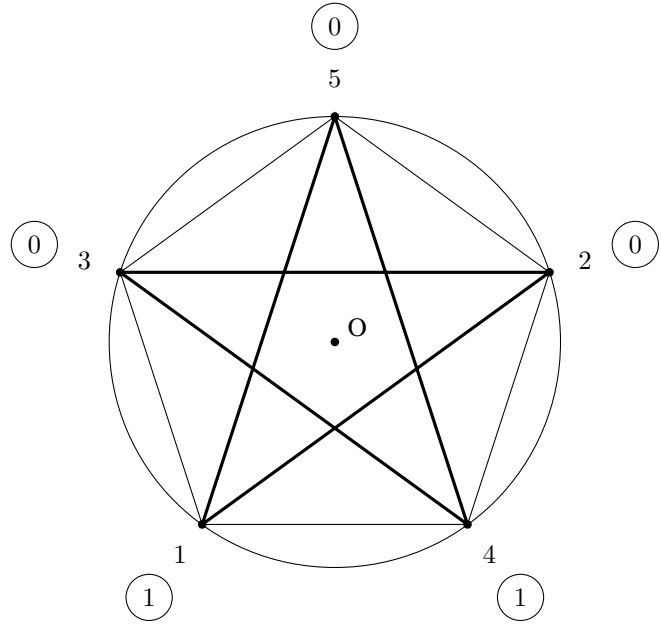

The probability of a successful simulation of the KS-box for random inputs $x = 1, \ldots, 5, y = 1, \ldots, 5$ is:
\begin{eqnarray}
\mbox{prob(successful sim)} & = & \frac{1}{25}(\sum_{x=y}p(a=b|x,y) \nonumber \\
& + & \sum_{\mbox{\small{p-gram edges}}}p(a\cdot b= 0|x, y) \nonumber \\
& + & \sum_{\mbox{\small{p-gon edges}}}p(a\cdot b = 0|x, y)) 
\end{eqnarray}
So for the two mixtures, $M_{1}$ and $M_{2}$:
\begin{eqnarray}
\mbox{prob(successful sim)}_{M_{1}} &= &\frac{1}{25} (5 + 10  + 10[1-(\frac{2}{3}\cdot \frac{1}{5} + \frac{1}{3} \cdot 0)]) \nonumber \\
& = & 1 - \frac{1}{25} \cdot \frac{4}{3}
\end{eqnarray}
\begin{eqnarray}
\mbox{prob(successful sim)}_{M_{2}} &= &\frac{1}{25} (5 + 10  + 10[1-(\frac{5}{6}\cdot \frac{1}{5} + \frac{1}{6} \cdot 0)]) \nonumber \\
& = & 1 - \frac{1}{25} \cdot \frac{5}{3}
\end{eqnarray}

Assuming the `$\perp$' constraint is satisfied for the pentagram edges, the first term in the sum refers to the five possible pairs of the same input for Alice and Bob, and the second term refers to the ten possible pairs of inputs corresponding to pentagram edges, where the probability of successful simulation is 1 in both cases. The third term refers to the ten possible pairs of inputs corresponding to pentagon edges, where the probability of failure is 1/5 in the case of $C_{2}$ charts and 0 in the case of $C_{1}$ or $C_{0}$ charts. So the \emph{optimal }probability of a successful simulation with classical resources is:
\begin{eqnarray}
\mbox{optimal prob(succesful sim)}_{C} & = & 1 - \frac{1}{25}\cdot\frac{4}{3} \nonumber \\
& \approx & .94667 \label{eq:optimal classical}
\end{eqnarray}

We now show that if Alice and Bob are allowed quantum resources, i.e., shared entangled states, they can achieve a greater probability of successful simulation of a 5-dimensional KS$_{p}$-box with $p = 1/3$ than the optimal classical strategy. 

First note that, analogously to the classical case, a perfect quantum simulation can be achieved for a marginal probability $p= 1/5$ if Alice and Bob initially (before separation) share copies of the maximally entangled state:
\begin{equation}
\frac{1}{\sqrt{5}} \sum_{i-1}^{5}\ket{i}\ket{i} \in \hil{H}_{5}\otimes\hil{H}_{5}  \label{eq:biorthog5}
\end{equation}
where $\{\ket{i}, i = 1, \ldots, 5\}$ is an orthogonal quintuple of states, i.e., a basis, in $\hil{H}_{5}$. The strategy is for Alice and Bob to produce outputs for given inputs $x = 1, \ldots, 5, y = 1, \ldots, 5$ via local measurements in this basis on their respective Hilbert spaces. The form of the biorthogonal representation with equal coefficients (\ref{eq:biorthog5}) guarantees that the outputs for the same inputs $x = y$ will satisfy the perfect correlation constraint (i) with $p=1/5$, and that the outputs for different inputs $x \neq j$ will satisfy the `$\perp$' constraint (ii)---which is simply an orthogonality constraint in this case---with $p=1/5$. For marginals $0 \leq p \leq 1/5$, Alice and Bob can mix this strategy for $p = 1/5$ with the strategy `output 0 for any input,' with the appropriate mixture probabilities, as in the classical case. 

For a marginal probability $p > 1/5$, a quantum simulation will have to adopt a different strategy. For the marginal $p = 1/3$,  suppose Alice and Bob initially (before separation) share many copies of the maximally entangled state
\begin{equation}
\frac{1}{\sqrt{3}} \sum_{i = 1}^{3} \ket{\alpha_{i}}\ket{\alpha_{i}} \in \hil{H}_{3}\otimes\hil{H}_{3}  \label{eq:biorthog3}
\end{equation} 
where $\{\ket{\alpha_{1}}, \ket{\alpha_{2}}, \ket{\alpha_{3}}\}$ is an orthogonal basis in $\hil{H}_{3}$. A biorthogonal representation with equal coefficients takes the same form for any basis. The strategy, for inputs $x = 1,\ldots,5, y = 1, \ldots, 5$, is for Alice to measure in \emph{any} basis containing the state $\ket{i}$ and for Bob to measure in \emph{any} basis containing the state $\ket{j}$, and to output the measurement outcome, where the states $\ket{i}$ and $\ket{j}$ are defined by the vertices of the pentagram/pentagon  on the circle $\Sigma_{2}$. The form of the biorthogonal decomposition (\ref{eq:biorthog3}) now guarantees that the outputs for the same inputs $x =y$ will be perfectly correlated, but  the outputs for any two different inputs $x \neq y$, will satisfy the `$\perp$' constraint for the \emph{pentagram} edges, which represent orthogonal pairs of states, but not for the \emph{pentagon} edges, which represent non-orthogonal pairs of states (as we have labeled the edges). 

The angle, $\chi$, subtended at $O$ by two non-orthogonal states corresponding to two radii of the unit sphere subtending an edge of the \emph{pentagon} (see Fig. 4) is given by:\footnote{This is the inverse of the golden ratio, the limit of the ratio of successive terms in the Fibonacci series: $\tau = \frac{\sqrt{5} + 1}{2}$: $1/\tau = \tau -1$.}

\begin{equation}
\cos \chi = \frac{\sqrt{5} - 1}{2}
\end{equation}
To see this, note that:
\begin{equation}
\sin \frac{\chi}{2} = s \sin \frac{\pi}{5} = \frac{\sin \frac{\pi}{5}}{\sqrt{2}\cos \frac{\pi}{10}} = 
\sqrt{2} \sin\frac{\pi}{10} = \sqrt{2}\frac{\sqrt{5}-1}{4}
\end{equation}

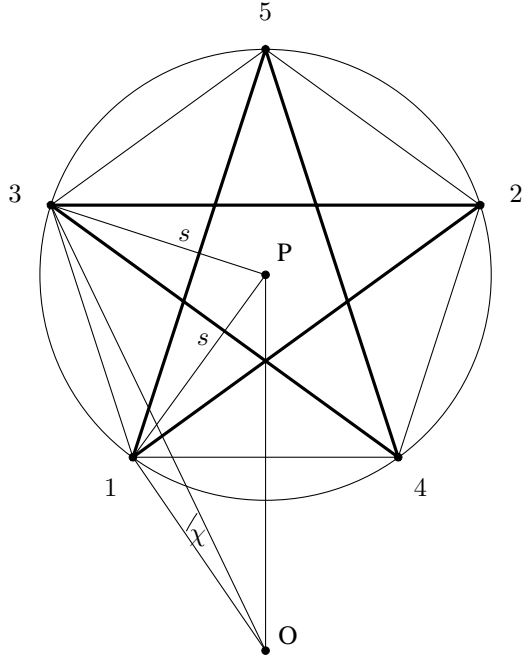
\begin{figure}[!ht]
    \begin{picture}(300,240)(-65,0)
\begin{tikzpicture}
\tikzstyle vertex=[circle,draw,fill=black,inner sep=1pt]
\path (0,0) coordinate (P);
\path (0,-5) coordinate (O);
\path (3*72+18:3cm) coordinate (P1);
\path (18:3cm) coordinate (P2);
\path (2*72+18:3cm) coordinate (P3);
\path (4*72+18:3cm) coordinate (P4);
\path (72+18:3cm) coordinate (P5);
\path (3*72+38:3.3cm) coordinate (Q);
\node at (.25,.3)  {P};
\node at (.3,-4.8)  {O};
\node at (-.9,-3.5)  {$\chi$};
\path (3*72+18:3.5cm)
node {$1$};
\path (3*72+10:1.2cm)
node {$s$};
\path (18:3.5cm) 
node {$2$};
\path (2*72+18:3.5cm) 
node{$3$};
\path (2*72+10:1.2cm) 
node{$s$};
\path (4*72+18:3.5cm)
node {$4$};
\path (72+18:3.5cm) 
node {$5$};
\draw[very thick] (P1) -- (P2) -- (P3) -- (P4) -- (P5) -- cycle;
\draw (P1) -- (P4) (P4) -- (P2)
(P2) -- (P5) (P5) -- (P3)
(P3) -- (P1);	
\draw (O) -- (P3);
\draw (O) -- (P1);
\draw (O) -- (P) (P) -- (P3) (P) -- (P1);
\draw (Q) arc (145:153:2cm);
\draw (0,0) circle (3cm);
\node[vertex] at (0,0) {};
\node[vertex] at (P1) {};
\node[vertex] at (P2) {};
\node[vertex] at (P3) {};
\node[vertex] at (P4) {};
\node[vertex] at (P5) {};
\node[vertex] at (O) {};

\end{tikzpicture}

\end{picture}
\caption{Pentagram  on $\Sigma_{2}$ showing angle $\chi$ between states $\ket{1}$ and $\ket{3}$}
    \label{fig4}
\end{figure}

It follows that the probability of success for a quantum simulation based on this strategy is given by:
\begin{eqnarray}
\mbox{prob(successful sim)}_{Q} & = & \frac{1}{25} (5 + 10  + 10[1-\frac{1}{3} (\frac{\sqrt{5} - 1}{2})^{2}]) \nonumber \\
& = & 1 - \frac{1}{25}\epsilon
\end{eqnarray}
where 
\[
\epsilon = 10(\frac{1}{3}(\frac{\sqrt{5}-1}{2}^{2}) \approx 10 \times .12732 < \frac{4}{3}
\]
i.e., a quantum simulation strategy based on shared maximally entangled states in  
$\hil{H}_{3}\otimes\hil{H}_{3}$ has a greater probability of success than the optimal classical strategy:
\begin{equation}
\mbox{prob(successful sim)}_{Q} \approx .94907 > \mbox{optimal prob(successful sim)}_{C} 
\end{equation}

\section{Simulating a PR-box with a KS-box}

As we have seen, a 5-dimensional KS$_{p}$-box with $p=1/3$ is nonclassical, so we expect the correlations to be monogamous. It is easy to see that they must be monogamous to avoid the possibility of signaling. 

For example, suppose Alice could share the KS-correlations with Bob and also with Charles. (We do not suppose that Bob and Charles share the KS-correlations.) Suppose Alice, Bob, and Charles all input 1. Then Bob's output must be the same as Charles' output, which means that:
\begin{equation}
p_{BC}(01|\mbox{Alice's input = 1}) = p_{BC}(10|\mbox{Alice's input = 1}) = 0
\end{equation}
where $p_{BC}(01|\mbox{Alice's input = 1}), p_{BC}(10|\mbox{Alice's input = 1})$ are the joint probabilities of different outputs for Bob and Charles, given that Alice inputs 1. Now suppose that Alice changes her input to 2. In this case, if Alice's output is 1 (which occurs with probability 1/3), Bob's output and Charles' output must both be 0. If Alice's output is 0 (which occurs with probability 2/3), Bob and Charles can jointly output 00 or 01 or 10 or 11, each with equal probability 1/6, i.e.,
\begin{equation}
p_{BC}(01|\mbox{Alice's input = 2}) = p_{BC}(10|\mbox{Alice's input = 2}) = 1/6
\end{equation}
So if Alice could share the KS-correlations with Bob and also with Charles, then Bob and Charles could detect the change in probability from 0 to 1/6 (the first measurement of a difference in their outputs would indicate this), and Alice could signal to Bob and Charles, i.e., `no signaling' entails monogamy. 

Consider, now, the problem of simulating a PR-box with a KS-box. That is, suppose Alice and Bob are equipped with 5-dimensional KS$_{p}$-boxes with $p = 1/3$ as communication channels. To what extent can they successfully simulate the correlations of a PR-box for random inputs 0 and 1? 

The following strategy has a probability of 3/4 for successful simulation: 
\begin{itemize}
\item Alice inputs 2 for PR-box input 0, and 1 for PR-box input 1
\item Bob inputs 3 for PR-box input 0, and 1 for PR-box input 1
\end{itemize}

To get the PR-box marginals of 1/2 for the outputs 0 and 1, Alice and Bob simultaneously flip their outputs randomly  for half the input pairs (i.e., before separation, they generate a random bit string, with equal probabilities for 0 and 1, which they share, and they associate successive rounds of the simulation---successive input pairs---with elements of the string; when the shared bit is 1, they both flip the output). Then:
\begin{itemize}
\item inputs 00 (i.e., KS-inputs 23) $\rightarrow$ outputs (00 or 11), 01, 10
\item inputs 01 (i.e., KS-inputs 21)  $\rightarrow$ outputs (00 or 11), 01, 10 
\item inputs 10 (i.e., KS-inputs 13)  $\rightarrow$ outputs (00 or 11), 01, 10 
\item inputs 11 (i.e., KS-inputs 11) $\rightarrow$ outputs 00, 11
\end{itemize}
with equal probability for each possibility, i.e., 1/3 for each of the outcomes (00 or 11), 01, 10 in the case of inputs 00, 01, 10,  and 1/2 for each of the outcomes 00, 11 in the case of inputs 11. 

If, in addition, Bob flips his output each round, then:
\begin{itemize}
\item inputs 00 (i.e., KS-inputs 23) $\rightarrow$ outputs (01 or 10), 00, 11
\item inputs 01 (i.e., KS-inputs 21)  $\rightarrow$ outputs (01 or 10), 00, 11 
\item inputs 10 (i.e., KS-inputs 13)  $\rightarrow$ outputs (01 or 10), 00, 11 
\item inputs 11 (i.e., KS-inputs 11) $\rightarrow$ outputs 01, 10
\end{itemize}

The (01 or 10) outputs for the input pairs 00, 01, 10 represent failures, and these occur with probability $3/4 \times 1/3 = 1/4$, so:
\[
\mbox{prob(successful sim)} = 3/4
\]

It is clear that there is no way of reducing the failure rate, so this is in fact the optimal strategy. It follows that a 5-dimensional KS$_{p}$-box with $p = 1/3$, which exhibits superquantum correlations, is classical with respect to nonlocality. That is, \emph{such a KS-box adds nothing to shared randomness as a resource in simulating the superquantum nonlocal correlations of a PR-box.}

This is confirmed by noting that, for any pair of inputs for Alice, and any pair of inputs for Bob, the CHSH inequality is satisfied by the correlations of a 5-dimensional KS$_{p}$-box with $p=1/3$, i.e., the maximum value of the correlation is equal to 2. 

To compare with the units in terms of which the CHSH inequality is usually expressed, where the observables take the values $\pm 1$, let $a = \pm 1$, $b = \pm 1$. Then for inputs $x = 1, \ldots, 5, y = 1, \ldots, 5$:
\begin{eqnarray}
\langle xy\rangle_{x=y}& = & 1 \\
\langle xy\rangle_{x \neq y} & = & -1/3
\end{eqnarray}
and for any $2 \times 2$ pairs of input values:
\begin{equation}
K = \langle xy\rangle + \langle xy'\rangle +\langle x'y\rangle - \langle x'y'\rangle \leq 2
\end{equation}
since at most two of these terms can be equal to 1, in which case the remaining two terms are each equal to -1/3.

It follows that the correlations for any two inputs for $x$ and any two inputs for $y$ can be recovered from a local hidden variable theory, but there is no product space that will generate the correlations between outputs for all possible input values to a 5-dimensional KS$_{p}$-box with $p = 1/3$, if the output values are required to be noncontextual, i.e., edge-independent (because the possibility of successfully simulating such a KS-box with only shared randomness as a resource is less than .95, as we saw in \S 3).

\section{Dropping the  Marginal Constraint}

For the marginal constraint:
\begin{equation}
p= 1/5
\end{equation}
we saw in \S 3 that a perfect classical simulation of a 5-dimensional KS-box  can be achieved with charts $C_{1}$. Similarly, a perfect quantum simulation can be achieved if Alice and Bob share copies of the maximally entangled state:
\begin{equation}
\frac{1}{\sqrt{5}} \sum_{i-1}^{5}\ket{i}\ket{i}
\end{equation}
where $\{\ket{i}, i = 1, \ldots, 5\}$ is a basis in $\hil{H}_{5}$.

If
\begin{equation}
0 \leq p \leq 1/5
\end{equation}
 a perfect simulation can be achieved if Alice and Bob  mix either of the above strategies with the strategy: `output 0 for any input,' with the appropriate mixture probabilities. 

If
\begin{equation}
1/5 \leq p \leq 1/3
\end{equation} 
Alice and Bob can mix the strategy for $p= 1/3$ in \S 3 with the strategy for $p = 1/5$, with appropriate mixture probabilities. A perfect simulation is impossible, but a quantum simulation is superior to the optimal classical simulation in this case, because a classical strategy will have to use  $C_{2}$ charts as well as $C_{1}$ and  $C_{0}$  charts .

For 
\begin{equation}
p = 1/2
\end{equation}
 if the inputs are the same, the outputs are required to be the same, with equal probability; if the inputs are different, then---since the output pair 11 has zero probability---it follows that the output pairs 01 and 10 must occur with equal probability 1/2 (i.e., the output pair 00 has zero probability). So, for this case, the correlations of a 5-dimensional KS-box become:
\begin{itemize}
\item if $x=y$, then $a=b$
\item if $x \neq y$, then $a \neq b$
\end{itemize}

It is now apparent that, for the marginal $p = 1/2$, and for pairs of inputs like $x \in \{1, 2\}, y \in \{1, 3\}$, i.e., where one of the inputs for Alice and Bob is the same and the other two different, \emph{a 5-dimensional  KS-box is equivalent to a PR-box.} If we interpret the KS-inputs $x = 2, 1$ as corresponding to the PR-inputs $x = 0,1$, respectively, and the KS-inputs $y = 3, 1$ as corresponding to the PR-inputs $y=0,1$, respectively, and Bob always flips his outputs, then the CHSH inequality is saturated and the correlations are precisely those of a PR-box, with the same marginals:
\begin{equation}
\langle 23\rangle + \langle 21\rangle + \langle 13\rangle - \langle 11\rangle = K_{PR} = 4
\end{equation}

If
\begin{equation}
1/3 \leq p \leq 1/2
\end{equation}
 a perfect simulation is impossible, but a quantum simulation is superior to a classical simulation. The CHSH inequality is violated: 
\begin{equation}
2 \leq K_{KS} \leq 4\, \mbox{when}\, 1/3 \leq p \leq 1/2
\end{equation}

The marginal probability $p =  \frac{1+\sqrt{2}}{6}$ yields the Tsirelson bound $2\sqrt{2}$. Note, however, that a perfect quantum simulation of all the correlations of  a 5-dimensional KS-box with this marginal is impossible, even though for any two inputs $x$ and any two inputs $y$, the KS-box is no more nonlocal than quantum mechanics---just as the correlations of the $p = 1/3$ case are superquantum, while being no more nonlocal than a classical theory for any two inputs $x$ and any two inputs $y$. 

As we noted in \S1, the space of `no signaling' bipartite probability distributions, with arbitrary inputs $x \in \{1, \ldots, n\}, y \in \{1, \ldots, n\}$ and binary outputs, 0 or 1 has the form of a convex polytope, with the vertices representing generalized PR-boxes (which differ only with respect to permutations of the inputs and/or outputs), or deterministic boxes, or (in the case $n > 2$) combinations of these. A 5-dimensional KS$_{p}$-box can be defined in terms of its joint probabilities as in Table 2. For $p=1/2$, the KSK-box is a generalized PR-box, with $p(00|xy) = p(11|xy) = 1/2$ in the diagonal cells, and $p(01|xy) = p(10|xy) = 1/2$ in the off-diagonal cells. Permuting the outputs for $y$ yields a box with  $p(01|xy) = p(10|xy) = 1/2$ in the diagonal cells, and $p(00|xy) = p(11|xy) = 1/2$ in the off-diagonal cells, in which case the probabilities for $x=2,1;  y = 3,1$ are as in the definition of a PR-box in \S1 (effectively a permutation of the inputs, with $x=2$ representing the PR-input $x=0$ and $y=3$ representing the PR-input $0$). It is now easy to see that the probabilities of a KS$_{p}$-box for $p < 1/2$ can be generated by mixing the extremal KS-box with $p = 1/2$ and the extremal deterministic box with $p(00|xy) = 1$ in each of the cells, in the ratio $2p: 1-2p$, so these KS-boxes lie inside the `no signaling' polytope.

\begin{table}[h!]
\begin{center}
\begin{tabular}{|ll||ll|ll|ll|lll|} \hline
   &$x$&$1$& &$2$ & &$\hdots$ & &$5$&&\\
   y&&&&&&&&&&\\
  \hline\hline $1$&&$1-p$ &$0$ &$1-2p$ &$p$ & $\ddots$ &&$1-2p$&$p$& \\
  &&$0$ &$p$ &$p$ &$0$ & & &$p$&$0$&\\\hline

 $2$&&$1-2p$ &$p$ &$1-p$ &$0$ & $\ddots$&  &$1-2p$&$p$&\\
  &&$p$ &$0$ &$0$ &$p$& &   &$p$&$0$&\\\hline

$\vdots$& & $\ddots$ &&  $\ddots$&&  $\ddots$ &&  $\ddots$&&\\\hline

  $5$&&$1-2p$ &$p$ &$1-2p$ &$p$ & $\ddots$&  &$1-p$&$0$&\\
  &&$p$ &$0$ &$p$ &$0$& &   &$0$&$p$&\\\hline

\hline
\end{tabular}
\end{center}
 \caption{Joint probabilities for a 5-dimensional KS$_{p}$-box.}
\end{table}

\section{Commentary}

An $n$-dimensional KS$_{p}$-box with marginal $p = 1/n$ can be perfectly simulated by a quantum simulation in which Alice and Bob share copies of the maximally entangled state $\frac{1}{\sqrt{n}} \sum\ket{i}\ket{i} \in \hil{H}_{n}\otimes\hil{H}_{n}$ and produce outputs for given inputs via local measurements in the same basis $\{\ket{i}, i = 1, \ldots, n\}$ on their respective Hilbert spaces, where the $n$ orthogonal basis states  are associated with the inputs $x = 1, \ldots, n, y = 1, \ldots, n$. The perfect correlation constraint (i) for the same inputs $x, y$ will be satisfied, and the `$\perp$' constraint (ii) for different inputs will be satisfied as a quantum orthogonality constraint. Similarly, a perfect classical simulation can be achieved if Alice and Bob share classical charts  with $n$ vertices selected by a shared random variable, in which a single vertex is pre-assigned the value 1 and the remaining $n-1$ vertices are pre-assigned the value 0. A perfect quantum or classical simulation can also be achieved for $0 \leq p \leq 1/n$ by mixing the strategy for $p = 1/n$ with the strategy `output 0 for any input' with the appropriate mixture probabilities, as we saw in \S 3 for the case $p = 1/5$. 

For $p > 1/n$, however, this is not possible, and a quantum simulation will have to adopt a strategy in which Alice and Bob produce outputs for given inputs on the basis of local measurements on copies of a shared entangled state $\frac{1}{\sqrt{m}} \sum_{i=1}^{m}\ket{i}\ket{i} \in \hil{H}_{m}\otimes\hil{H}_{m}$ with $m < n$ to generate the marginal probability. Then different input pairs $x,y$; $x',y'$ can be associated with different local measurement contexts defined by different bases in $\hil{H}_{m}$, and it is possible that the same input can be associated with two or more \emph{incompatible} local measurement contexts (as we saw for $p = 1/5$ in \S 3, where the input corresponding to a vertex of the pentagram could be associated with two contexts associated with two bases in $\hil{H}_{3}$ represented by the two edges of the pentagram intersecting in the vertex). A little reflection shows that if each input can be associated with two or more incompatible local measurement contexts associated with different bases, then $n \geq 5$. 

If $n=2$,  a perfect quantum simulation is possible for all marginal probabilities $0 \leq p \leq 1/2$ if Alice and Bob share copies of the maximally entangled state $\frac{1}{\sqrt{2}} \sum\ket{i}\ket{i}$ in $\hil{H}_{2}$. There can only be one local measurement context associated with each input, because the state $\ket{i} \in \hil{H}_{2}$ corresponding to the input $i$ cannot belong to two different bases in $\hil{H}_{2}$. 

If $n=3$, there are three possible local measurement contexts represented by the input pairs 12, 13; 21, 23; 31, 32 (we take permutations of  contexts such as 12 and 21 as equivalent). The orthogonality relations of the three contexts are represented by the edges of a triangle, in which each vertex (corresponding to an input of the KS-box) is associated with two contexts. Clearly, in $\hil{H}_{3}$, the three contexts can be embedded into a single context associated with a basis in $\hil{H}_{3}$ (since the triangle also represents  the orthogonality relations of the three basis states).   In order for each input to be associated with two incompatible local contexts in a quantum simulation, the two contexts would have to be represented by orthogonal bases with a common basis state in a proper subspace of $\hil{H}_{3}$, i.e., in a 2-dimensional Hilbert space, which is impossible. 

If $n=4$, there are six possible local measurement contexts represented by the input pairs 12, 13, 14; 21, 23, 24; 31, 32, 34; 41, 42, 43. The orthogonality relations of the six contexts are represented by the edges and diagonals of a square, in which each vertex is associated with three contexts. Again, the six contexts can be embedded into a single context associated with a basis in $\hil{H}_{4}$ (since the square with diagonals also represents the orthogonality relations of the four basis states). In order for each input to be associated with at least two incompatible local contexts in a quantum simulation, the two contexts would have to be represented by different bases with a common basis state in a proper subspace of $\hil{H}_{4}$, i.e., in a 3-dimensional Hilbert space (since this is impossible on $\hil{H}_{2}$).  If we remove two edges of the square with diagonals, in such a way that each vertex is associated with two contexts, the orthogonality relations in $\hil{H}_{3}$ are inconsistent with the assumption that there are four distinct vertices, each associated with two incompatible local contexts.

 \begin{figure}[!ht]
    \begin{picture}(300,80)(-15,0)
\begin{tikzpicture}
\tikzstyle vertex=[circle,draw,fill=black,inner sep=1pt]
\path (0,0) coordinate (O);
\path (3*72+18:1cm) coordinate (P1);
\path (18:1cm) coordinate (P2);
\path (2*72+18:1cm) coordinate (P3);
\path (4*72+18:1cm) coordinate (P4);
\path (72+18:1cm) coordinate (P5);
\path (-4.4,-.8) coordinate (S1);
\path (-2.5,-.8) coordinate (S2);
\path (-2.5,.9) coordinate (S3);
\path (-4.4,.9) coordinate (S4);
\path (-8,-.8) coordinate (T1);
\path (-6,-.8) coordinate (T2);
\path (-7,.9) coordinate (T3);
\path (-9.6,-.8) coordinate (L1);
\path (-9.6,.9) coordinate (L2);
\path (-9.6,-1.1)
node {$1$};
\path (-9.6,1.2)
node {$2$};
\path  (-8,-1.1)
node {$1$};
\path  (-6,-1.1)
node {$2$};
\path  (-7,1.2)
node {$3$};
\path (-4.4,-1.1)
node {$1$};
\path (-2.5,-1.1)
node {$2$};
\path (-2.5,1.2)
node {$3$};
\path (-4.4,1.2)
node {$4$};
\path (3*72+18:1.3cm)
node {$1$};
\path (18:1.3cm) 
node {$2$};
\path (2*72+18:1.3cm) 
node {$3$};
\path (4*72+18:1.3cm)
node {$4$};
\path (72+18:1.3cm) 
node {$5$};
\draw (S1) -- (S2) -- (S3) -- (S4) -- cycle;
\draw (S1) -- (S3);
\draw (S2) -- (S4);
\draw (T1) -- (T2) -- (T3) -- cycle;
\draw (L1) -- (L2);
\draw (P1) -- (P2) -- (P3) -- (P4) -- (P5) -- cycle;
\draw (P1) -- (P4) (P4) -- (P2)
(P2) -- (P5) (P5) -- (P3)
(P3) -- (P1);	
\node[vertex] at (L1) {};
\node[vertex] at (L2) {};
\node[vertex] at (T1) {};
\node[vertex] at (T2) {};
\node[vertex] at (T3) {};
\node[vertex] at (S1) {};
\node[vertex] at (S2) {};
\node[vertex] at (S3) {};
\node[vertex] at (S4) {};
\node[vertex] at (P1) {};
\node[vertex] at (P2) {};
\node[vertex] at (P3) {};
\node[vertex] at (P4) {};
\node[vertex] at (P5) {};
\end{tikzpicture}
    \label{fig5}
\end{picture}
\caption{Basis orthogonality relations in $\hil{H}_{n}$,  for $n= 2, 3, 4, 5$}
 
\end{figure}
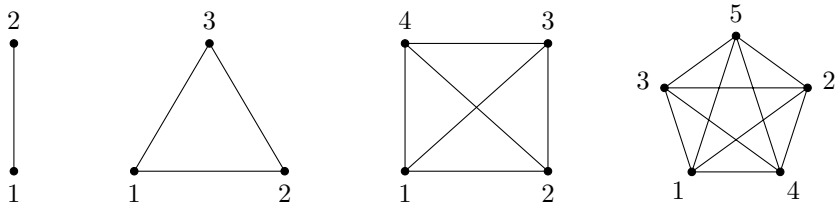

For example, suppose we remove the two diagonals of the square. The vertices 1 and 3 are represented by 1-dimensional projection operators in $\hil{H}_{3}$ that are both orthogonal to the plane defined by the 1-dimensional projectors representing vertices  2 and 4, which requires that 1 and 3 are represented by the same 1-dimensional projector. (See Fig. 5.)

If $n=5$, there are ten possible local measurement contexts. The orthogonality relations are represented by the edges and diagonals of a pentagon, i.e., a pentagon with an inscribed pentagram, in which each vertex is associated with four contexts. The ten contexts can be embedded into a single context associated with a basis in $\hil{H}_{5}$ (since the pentagon and pentagram edges also represent the orthogonality relations of the five basis states). If we remove the diagonals (the edges of the pentagram), or if we remove the edges of the pentagon, each vertex is associated with two contexts. As we showed in \S 3, the orthogonality relations of the pentagram (or, equivalently, the pentagon, but not both) can be implemented in $\hil{H}_{3}$, in such a way that each vertex is associated with two incompatible local contexts. 

Note that in considering a classical or quantum simulation of an $n$-dimensional KS$_{p}$-box with $p > 1/n$, there is a trade-off between satisfying the marginal constraint and perfectly simulating the correlations. For example, in the case of the 5-dimensional KS$_{p}$-box with $p = 1/3$, one could always adopt a strategy for simulating the $p = 1/3$ case by mixing the strategy for $p = 1/5$ with the classical or quantum strategy for $p = 1/3$ considered in \S 3. The simulation will fail on two counts: with respect to meeting the marginal constraint, and with respect to recovering the correlations. But such a strategy will do better at recovering the correlations than the strategy for $p = 1/3$, and will also achieve a marginal probability for the output 1 that is closer to the value $p = 1/3$ than the strategy for $p = 1/5$. What is clear, though, is that the closer a simulation strategy approximates the correlation constraint, the more the value of $p$ decreases from the required value of 1/3 (where the probability of meeting the correlation constraint is less than .95) to 1/5 (where the probability of meeting the correlation constraint is 1). In the preceding discussion, we opted to consider the question of simulating the correlations of an $n$-dimensional KS$_{p}$-box under the assumption that the simulation meets the marginal constraint.

The space of `no signaling' bipartite theories for two binary-valued observables for each party---equivalently the space of `no signaling' bipartite probability distributions with binary-valued inputs and binary-valued outputs---can be divided  into a classical region bounded by the value 2 for the CHSH correlation $K = \langle 00\rangle + \langle 01\rangle + \langle 10\rangle - \langle 11\rangle$, a quantum region bounded by the Tsirelson bound $2\sqrt{2}$, and a superquantum region between the Tsirelson bound and the maximum value $K = 4$ attained by a PR-box:  
\begin{eqnarray}
K_{C} & \leq & 2  \label{eq:CHSH(C)} \\
K_{Q} & \leq & 2\sqrt{2} \label{eq:CHSH(Q)} \\
K_{PR} & = & 4 \label{eq:CHSH(PR)}
\end{eqnarray}
Any probability distribution in the classical region can be represented as a unique mixture (convex combination) of bipartite pure states that are locally deterministic for each party, represented by vertices of the classical polytope, which is a simplex. It follows that the distribution can be generated by a random variable shared between the two parties, where the values label local deterministic states assigning values to given inputs. Probability distributions in the region outside the classical simplex exhibit correlations that are more nonlocal than classical correlations. Each probability distribution in the nonclassical region can be represented non-uniquely as a mixture of pure or extremal states represented by vertices of the `no signaling' polytope, which is not a simplex.

A KS-box, as a hypothetical superquantum information channel, reveals a further dimension of structure in the information-theoretic properties of `no signaling' theories, having to do with the contextuality of theories outside the classical simplex. To reveal this structure requires considering theories with more than two observables for each party.

Consider a 5-dimensional KS-box with $p=1/3$. Referring to the discussion in \S 3, let
\begin{equation}
Z =  |\sum_{\mbox{\small{p-gram edges}}}p(a\cdot b= 0|x, y)  - \sum_{\mbox{\small{p-gon edges}}}p(a\cdot b = 0|x, y))| 
\end{equation}
and define the correlation:
\begin{equation}
\mathcal{K}_{C} = \sum_{x=y}p(a=b|x,y) - Z 
\end{equation}
It follows from equation (\ref{eq:optimal classical}) in \S 3 that 
the optimal classical value for $\mathcal{K}$ is:
\begin{equation}
\mathcal{K_{C}} = 5 - \frac{4}{3}
\end{equation}
This expresses a constraint on the probabilities derived from a noncontextual assignment of 0's and 1's to the inputs $1,\ldots,5$ satisfying the orthogonality constraint, either for the pentagram edges or for the pentagon edges, where local noncontextuality in satisfying the orthogonality constraint is forced by the requirement of perfect correlation for the same inputs $x=y$. For a 5-dimensional KS-box, we have:
\begin{equation}
\mathcal{K}_{KS}=5 
\end{equation}

Since $\mathcal{K}_{C} < \mathcal{K}_{KS}$, the correlations of a 5-dimensional KS-box with $p = 1/3$ cannot be recovered from a probability distribution that lies inside the classical simplex.
However, for any subset of  $2 	\times 2$ input pairs $K < 2$, so the correlations for any particular subset of $2 \times 2$  input pairs can be recovered from a probability distribution that lies inside the classical simplex. In other words, if Alice and Bob are told in advance that they will be required to simulate the correlations of  a particular subset of $2 \times 2$ input pairs to a 5-dimensional KS-box with $p = 1/3$, there is a local strategy based on shared randomness that will enable them to do so. What is significant here is that the different classical  local `contexts' defined by the classical simplices associated with the different subsets of $2 \times 2$ input pairs cannot be embedded into the classical simplex for all 5 x 5 input pairs. Note that the lattice of subspaces of a simplex is a Boolean algebra, with a 1-1 correspondence between the vertices and the facets (the $(n-1)$-dimensional faces). So the `contexts' defined by these classical simplices are Boolean algebras.

For the maximally entangled quantum state in $\hil{H}_{3}\otimes\hil{H}_{3}$, we obtain:
\begin{equation}
\sum_{x=y}p(a=b|x,y) - Z =5 - \epsilon
\end{equation}
where $\epsilon < \frac{4}{3}$. 
We conjecture that this is the optimal quantum value $\mathcal{K_{Q}}$.
The fact that the quantum bound exceeds the classical bound reflects a feature of quantum probability assignments that is not shared by classical probability assignments: not only is the `$\perp$' constraint satisfied as an orthogonality constraint  by 0, 1 probabilities for the orthogonal pentagram edges, but the `$\perp$' constraint is satisfied for the non-orthogonal pentagon edges probabilistically, in the sense that the probability that two non-orthogonal vertices are both assigned the value 1 decreases continuously with the square of the cosine of the angle between the vertices, as the angle varies between 0 and orthogonality.

The inequality
\begin{equation}
\mathcal{K}_{C} < \mathcal{K}_{Q} < \mathcal{K}_{KS}
\end{equation}
then expresses the relative extent to which the correlations of each type of theory are contextual, in the sense that the correlations for all inputs (or all observables) cannot be derived from a joint probability distribution for all pairs of inputs in the classical simplex, even though the correlations for every subset of $2 \times 2$ inputs can be derived from a joint probability distribution that lies inside the corresponding classical simplex---i.e., these classical local `contexts' cannot be embedded into the classical simplex for the full set of joint probabilities.  A KS-box can be superquantum with respect to contextuality as measured by the correlation $\mathcal{K}$, while being no more nonlocal than a classical theory, as measured by the CHSH correlation $K$ for any subset of  $2 \times 2$ input pairs. Similarly, since the Tsirelson bound can be attained by the correlations for certain subsets of  $2 \times 2$ input pairs to a 5-dimensional KS$_{p}$-box with $p > 1/3$, while a perfect quantum simulation for all pairs of inputs is impossible, it follows that  a KS-box can be superquantum with respect to contextuality, as measured by the  correlation $\mathcal{K}$, while being no more nonlocal than quantum mechanics, as measured by the CHSH correlation $K$ for any subset of  $2 \times 2$ input pairs.

\section*{Acknowledgements}

Jeffrey Bub acknowledges support from  the University of Maryland Institute for Physical Science and Technology and informative discussions with Daniel Rohrlich. Allen Stairs acknowledges support from NSF grant no. 0822545.

\bibliographystyle{plain}
\bibliography{ks}

\end{document}